\documentclass{article} 

\usepackage{preprint}

\usepackage[utf8]{inputenc}	
\usepackage[T1]{fontenc}	
\usepackage{xcolor}		
\usepackage[colorlinks = true,
            linkcolor = purple,
            urlcolor  = blue,
            citecolor = cyan,
            anchorcolor = black]{hyperref}	
\usepackage{nicefrac}		
\usepackage{microtype}		
\usepackage{float}			
\usepackage{multicol}		

\usepackage{newfloat}
\DeclareFloatingEnvironment[name={Supplementary Figure}]{suppfigure}
\usepackage{sidecap}
\sidecaptionvpos{figure}{c}

\usepackage{titlesec}
\titlespacing\section{0pt}{12pt plus 3pt minus 3pt}{1pt plus 1pt minus 1pt}
\titlespacing\subsection{0pt}{10pt plus 3pt minus 3pt}{1pt plus 1pt minus 1pt}
\titlespacing\subsubsection{0pt}{8pt plus 3pt minus 3pt}{1pt plus 1pt minus 1pt}

\ifpdf
  \pdfoutput=1\relax                   
  \usepackage{graphicx}                
  \DeclareGraphicsExtensions{.pdf,.png,.jpg,.jpeg} 
\else
  \usepackage{graphicx}                
  \DeclareGraphicsExtensions{.eps}     
\fi%

\PassOptionsToPackage{warn}{textcomp}  
\usepackage{textcomp}                  
\usepackage{mathptmx}                  
\usepackage{times}                     
\usepackage{cite}                      
\usepackage{tabu}                      
\usepackage{booktabs}                  

\usepackage{soul}
\usepackage{balance}

\title{SIGMA: An Open-Source Interactive System \\for Mixed-Reality Task Assistance Research}

\usepackage{authblk}

\author{
 Dan Bohus$^1$, Sean Andrist$^1$, Nick Saw$^1$, Ann Paradiso$^1$, Ishani Chakraborty$^2$, Mahdi Rad$^2$ \\
 \texttt{\{dbohus;sandrist;chitsaw;annpar;ischakra;mahdirad\}@microsoft.com}\\
 $^1$Microsoft Research\\
 $^2$Microsoft
}

\begin{document}


\maketitle

\begin{figure*}[h]
 \centering 
 \includegraphics[width=\textwidth]{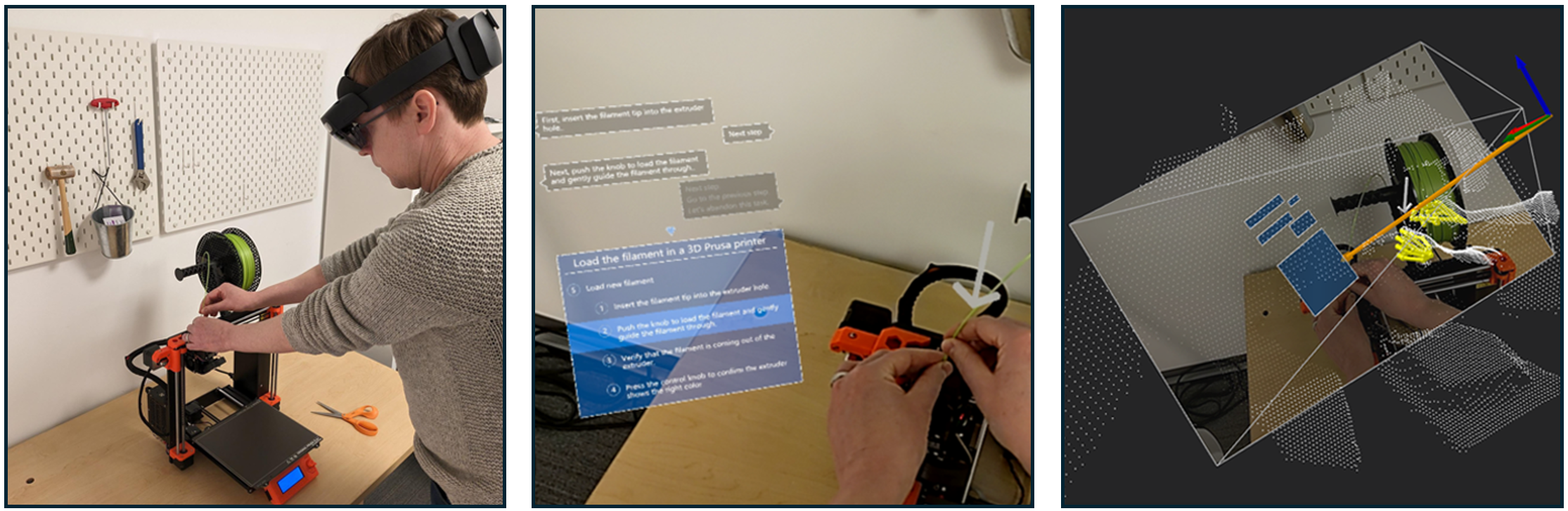}
 \caption{\textbf{Left}: user performing a procedural task with a mixed-reality headset running \textsc{Sigma}. \textbf{Middle}: first-person view showing \textsc{Sigma} guidance panel and task-specific holograms. \textbf{Right}: visualization of system's scene understanding showing the egocentric camera view, depth map, detected objects, gaze, hand and head pose in 3D space. \textcopyright \emph{2024 IEEE}}
 \label{fig:teaser}
\end{figure*}

\begin{abstract}
We introduce an open-source system called \textsc{Sigma} (short for ``Situated Interactive Guidance, Monitoring, and Assistance'') as a platform for conducting research on task-assistive agents in mixed-reality scenarios. The system leverages the sensing and rendering affordances of a head-mounted mixed-reality device in conjunction with large language and vision models to guide users step by step through procedural tasks. We present the system's core capabilities, discuss its overall design and implementation, and outline directions for future research enabled by the system. \textsc{Sigma} is easily extensible and provides a useful basis for future research at the intersection of mixed reality and AI. By open-sourcing an end-to-end implementation, we aim to lower the barrier to entry, accelerate research in this space, and chart a path towards community-driven end-to-end evaluation of large language, vision, and multimodal models in the context of real-world interactive applications.
\end{abstract}
\vspace{0.35cm}


\begin{multicols}{2} 


\section{Introduction}
Enabling fluid human-machine collaboration in the physical world has been a long-standing pursuit in the AI research community. Building useful collaborative systems requires weaving together several key competencies, from sensing and reasoning about the user and their surrounding physical context, to taking action and generating spoken utterances with precise timing and coordination in the flow of activity. Intelligent collaborative systems have the potential to support people to accomplish complex tasks under stress or pressure, help them learn new skills, correct and anticipate errors, and generally augment human capabilities to make people more self-sufficient in their everyday lives.

Many robotics technologies have been envisaged and developed in this space, aiming to produce systems that can contribute to a collaboration by taking physical actions. Augmented or mixed-reality technologies provide a different, yet powerful conduit for empowering individuals by providing just-in-time assistance in the physical world. By generating contextualized, world-aligned graphics and instructions, mixed-reality-based task assistants can bridge the virtual and physical worlds and match users' goals, needs, and preferences. Unsurprisingly, procedural task assistance has been one of the earliest envisioned applications for mixed reality \cite{early} and has continued to attract significant attention \cite{bohus2005larri, darpa_ptg, zheng2015eye, malta2021augmented, puladi2022augmented, castelo2023argus}. Developing such systems raises challenges and opportunities across a host of interconnected research areas, including computer vision, AI, machine learning, and human-computer interaction.

Recent breakthroughs in generative AI and advances in large language, vision, and multimodal models (\emph{e.g.}, GPT-4 \cite{gpt4}, CLIP \cite{clip}, Detic \cite{detic}, LLaMA \cite{llama}, LLaVA \cite{llava}, Kosmos \cite{kosmos}, OpenFlamingo \cite{openflamingo}) have opened up new possibilities in this space. By providing a substrate of open-domain knowledge, inference, and generation capabilities, these models have the potential to enable task assistance in open-ended (rather than predefined) scenarios. However, these new technologies also raise new challenges around how to utilize them effectively in zero-shot or few-shot regimes; how to integrate into real-time, streaming, egocentric settings; how to improve robustness, reduce hallucinations, and mitigate safety risks; and so on. While evaluations on benchmark datasets can be informative, the true test for the usefulness of these models is in the context of complete end-to-end systems that interact with users. However, given their complex, multimodal nature, developing complete systems requires a significant amount of infrastructure, engineering, and maintenance effort. More shared tools and open systems are needed to make faster progress.

In this paper we introduce \textbf{\textsc{Sigma}}, shorthand for \textbf{S}ituated \textbf{I}nteractive \textbf{G}uidance, \textbf{M}onitoring and \textbf{A}ssistance, an open-source\footnote{\url{https://aka.ms/psi-sigma}} platform for conducting research in mixed-reality task assistance. \textsc{Sigma} provides an initial system implementation that leverages large language and large vision models to guide users through procedural tasks. The system relies on a client-server architecture where sensor capture and rendering is performed on a HoloLens 2 device, but perception and computation is offloaded live to a compute server, allowing researchers to bypass computational limits on current headset devices, and fostering device-independent research and development. By open-sourcing the system, we aim to lower the barrier to entry and accelerate research at the intersection of mixed reality and AI. We believe a common platform can serve as a basis for more reproducible research, and in the longer run allows for constructing interactive, user-in-the-loop challenges and evaluations for various inferential and generative AI technologies.

\section{Related Work}

Providing in-stream assistance in physical tasks has long been identified as one of the most promising application scenarios for augmented reality technologies. Important opportunities created by the ability to overlay world-aligned graphics and instructions in a user's field of view have been recognized by Caudell and Mitzell \cite{early}, who demonstrated an early system for wiring, layout and assembly tasks.

Since that early work, various research efforts have explored a diverse set of connected application areas, including machine maintenance and repair \cite{bohus2005larri, zheng2015eye, malta2021augmented, henderson_repair}, assembly tasks \cite{tang2003comparative}, training and education \cite{liu2018smart, webel2013augmented}, surgery and medical applications \cite{puladi2022augmented, jiang2020hololens}, etc. Many important questions have been brought to the fore and investigated over time, such as challenges with dialog management and with preparing maintenance manuals for use in a language interface \cite{bohus2005larri}; human-factors implications for designing more effective interactions using AR technologies  \cite{zheng2015eye, tang2003comparative}; techniques for adaptively projecting content to augment the abilities of neurodiverse and cognitively impaired workers \cite{funk2015using}; spatial localization requirements for AR-assisted precise surgical procedures \cite{jiang2020hololens}; and so on.

Recently, the DARPA Perceptually-enabled Task Guidance (PTG) program has aimed to advance the state-of-the-art in mixed-reality task assistance by developing AI technologies that help users perform complex physical tasks, while reducing errors and expanding their skillset \cite{darpa_ptg}. \textsc{Tim}, short for Transparent, Interpretable and Multimodal AR Personal Assistant \cite{castelo2023argus}, is a mixed-reality task assistive system developed under the aegis of DARPA PTG. Like \textsc{Sigma}, the system uses a HoloLens 2 device in a client-server architecture, and comes with extensive tools for data visualization and debugging.

Several datasets capturing egocentric perspectives have appeared in recent years \cite{epickitchens, bao2023foundation, ego4d, grauman2023egoexo4d, assembly101, egoprocel, holoassist}---an important resource for training machine learning models for mixed-reality task assistants. For example, Epic Kitchens \cite{epickitchens} provides a large-scale collection of first-person activities in the kitchen. Ego4D \cite{ego4d} provides over 3600 hours of egocentric daily life activity data collected at over 74 locations spanning 9 countries. Ego-Exo4D \cite{grauman2023egoexo4d} is an evolution of that dataset, with 1286.3 hours of video of skilled human activities, captured from both egocentric and exocentric sensors. Assembly101 \cite{assembly101} focuses on assembly and disassembly tasks and provides simultaneous static and egocentric recordings, while EgoProceL \cite{egoprocel} focuses on multimodal procedural learning from egocentric views. Most closely related to the mixed-reality procedural task assistance scenario targeted by \textsc{Sigma}, HoloAssist \cite{holoassist} is a large-scale egocentric human interaction dataset in which a performer using a HoloLens 2 headset is assisted by a human instructor to complete various tasks in the physical world. Finally, "Watch, Talk and Guide" (WTaG) \cite{bao2023foundation} is a recent multimodal benchmark dataset intended to explore the application of state-of-the-art foundation models for situated interactive task guidance.

Beyond basic research, datasets, and demonstration systems, commercial solutions for mixed-reality task assistance have also been constructed and deployed in real-world use cases. Microsoft's Dynamics 365 Guides \cite{guides} is a HoloLens 2 based system that provides holographic step-by-step instructions in industrial settings, coupled with task guide authoring, and remote assistance features.

A number of open-source simulation platforms have also emerged to provide support for studying embodied AI tasks. For example, Alexa Arena \cite{alexaarena} offers diverse multi-room layouts and interactable objects. AI2-THOR \cite{ai2thor} features nearly photo-realistic 3D indoor scenes tailored for tasks related to navigation and interaction. ALFRED \cite{alfred} is an action-centric benchmark that focuses on the interpretation of grounded instructions within everyday scenarios.

Altogether, these resources create important opportunities for advancing the state of the art in mixed-reality task assistance. Even so, many interesting research questions can only be properly investigated in the context of end-to-end systems that are deployed live and interact with real users. Developing such systems requires a significant amount of engineering expertise and effort. 

\begin{figure*}[t!]
 \centering 
 \includegraphics[width=\textwidth]{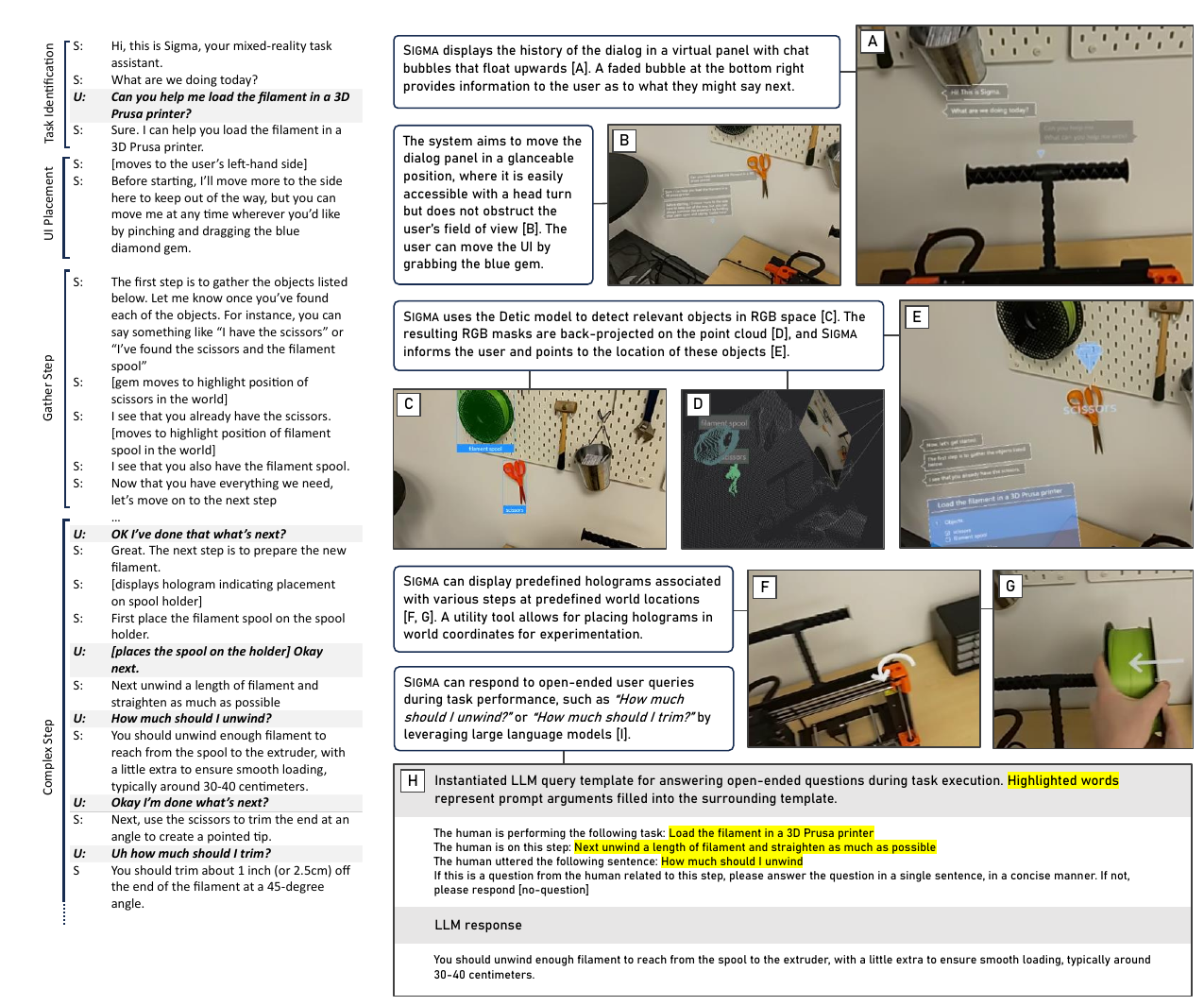}
 \caption{An illustrative example of an interaction with the \textsc{Sigma} system. (portions \textcopyright 2024 IEEE)}
 \label{fig:example}
\end{figure*}

\textsc{Sigma} aims to address this gap. Our primary goal and contribution is to provide an open, extensible, robust platform for mixed-reality task assistance research. \textsc{Sigma} builds upon a modern infrastructure for working with multimodal streaming data \cite{psi}, and provides an initial set of features that leverage large language \cite{azure_openai, gpt4} and large vision models \cite{detic, seem} for interactive task guidance. To our knowledge, \textsc{Sigma} is the first open-source system of this kind.

\section{System Functionality}

We begin by introducing \textsc{Sigma}'s core functionality via a sample interaction, illustrated in Figure \ref{fig:example}.

\textsc{Sigma} can engage in spoken dialog with the user, supported by virtual displays and holograms. Throughout the interaction, the system displays the history of the dialog in a set of chat bubbles, with system utterances on the left-hand side and user utterances on the right-hand side, as shown in Figure \ref{fig:example}.A, B and E. Taking inspiration from the \textsc{DiamondHelp} system \cite{diamondhelp}, when it is the user's turn to speak, \textsc{Sigma} also displays a chat bubble on the bottom right with faded text containing a subset of example utterances the user might say next (Figure \ref{fig:example}.A).

Interactions begin with a task identification sub-dialog where the system starts by asking the user which task they need assistance with. The system can guide the user through tasks in two modes. In the first mode, preexisting tasks are manually defined and specified in a task library in simple \texttt{.json} format. \textsc{Sigma} uses a large language model to perform intent recognition and map the user's request to one of the preexisting tasks. Alternatively, the second mode allows for automatically generating the entire recipe for any desired task using LLM queries. In this mode, the system first uses an LLM query to generate a set of questions that elicit more information about the user's context. After receiving the user's responses to these questions, the system uses another LLM query to generate contextualized step-by-step instructions for the task at hand.

Once the task recipe is identified (or generated via the LLM), \textsc{Sigma} begins to guide the user through the task. Step-by-step instructions are displayed in a virtual task panel placed below the chat bubbles, with the current step highlighted, as shown in Figure \ref{fig:example}.E. The application currently supports linear tasks (no-branching) containing three types of steps: a \emph{simple} step providing a single instruction to the user, a \emph{complex} step that provides a high level instruction decomposing into a linear set of sub-steps (each with their own instructions), and a \emph{gather} step that instructs the user to find and collect a set of objects required for performing the task, such as tools, ingredients, etc.

In the example from Figure \ref{fig:example}, the task recipe is retrieved from the predefined task library, and the first step is a \emph{gather} step. Throughout the interaction, \textsc{Sigma} can be configured to use an open-world object detection model \cite{detic} to detect and track the set of task-relevant objects (additional implementation details are presented in section \ref{sec:object_tracking}). If the system finds one of the relevant objects in the user's environment, it draws the user's attention to the object's location with a hologram animation and a virtual label with the object's name above its location, as shown in Figure \ref{fig:example}.E. The interaction during the gather step is mixed-initiative \cite{mixed_initiative}: the system calls out objects automatically as it finds them, but the user can also specify that certain objects are available if they are not automatically detected. Once all the relevant objects have been identified, the system moves on to the next step in the recipe.

For \emph{simple} steps, \textsc{Sigma} speaks and displays the corresponding step instruction to the user and waits for the user to report completion of the step. For \emph{complex} steps, \textsc{Sigma} first presents the high-level instruction, then each sub-step one-by-one (waiting for the user to report completion of each sub-step). At each step or sub-step, the system may also display holograms, such as a straight arrow pointing to a button to be pressed, a curved arrow indicating a motion path for twisting a knob, custom holograms to support specific actions in the task, etc. (Figure \ref{fig:example}.F and G).

During the execution of \emph{simple} and \emph{complex} steps, \textsc{Sigma} leverages an LLM to answer open-ended questions from the user. This feature is accomplished with another LLM query, illustrated in Figure \ref{fig:example}.H, which has been designed to identify whether a user utterance is a question (e.g., as opposed to self-talk) and if so, generate a corresponding answer based on the context of the task and the current step. The resulting answer is displayed and spoken back to the user just like any other system utterance.

\begin{figure}[H]
 \centering 
 \includegraphics[width=\columnwidth]{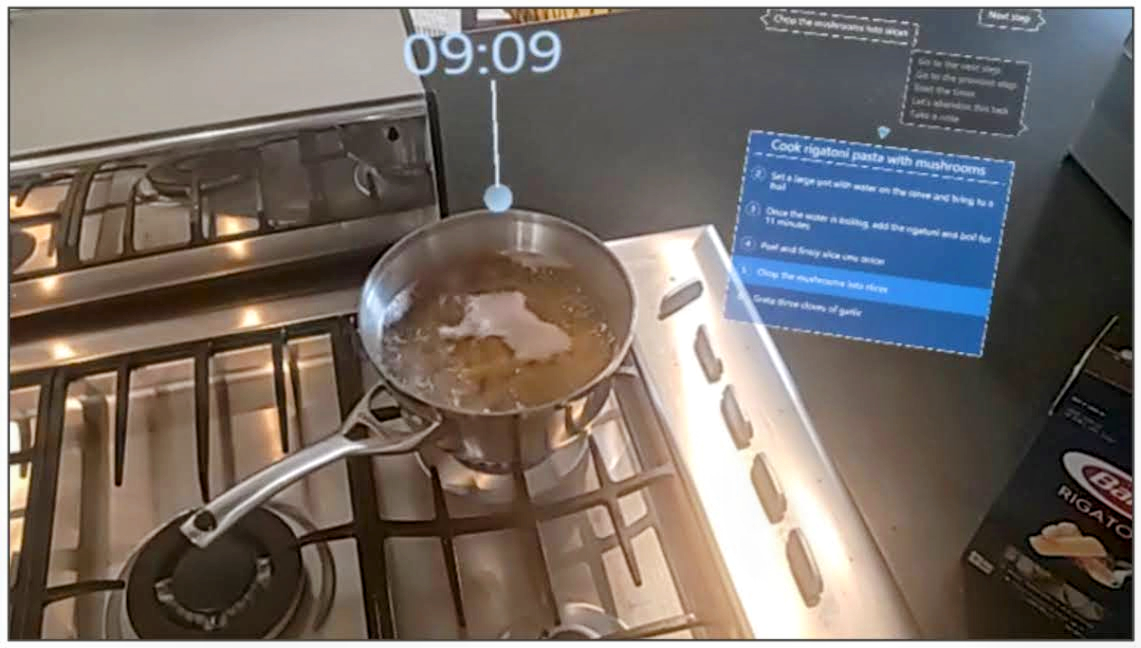}
 \caption{First-person view during a cooking task showing a spatially-placed timer hologram.}
 \label{fig:cooking}
\end{figure}

Another system feature is the ability to set \emph{countdown timers}. Individual steps from a task in the library may optionally define an expected duration. In this case, during the interaction, the user can start a countdown timer with a simple voice command. A virtual timer is displayed in front of the user and the user can move it to any desired location by grabbing and moving a handle. Spatially placed timers can be helpful in a variety of tasks where certain steps have an expected duration, for instance during cooking, as illustrated in Figure \ref{fig:cooking}.

The set of features described above and available in the initial version of the system constitute a starting point for developing and studying more sophisticated guidance, monitoring, and assistance interactions. These features are by no means comprehensive---rather, the system is intended to serve as a starting point, and is easily extensible to accommodate myriad research investigations and scenarios. We review possible extensions and research directions in Section \ref{sec:Discussion}.

\section{System Implementation}

We now turn our attention to implementation aspects and discuss several important details and affordances for supporting prototyping and research. We begin with an architectural overview.

\subsection{Architectural Overview}
\label{subsec:ArchitecturalOverview}

\begin{figure*}[tb]
 \centering 
 \includegraphics[width=1.5\columnwidth]{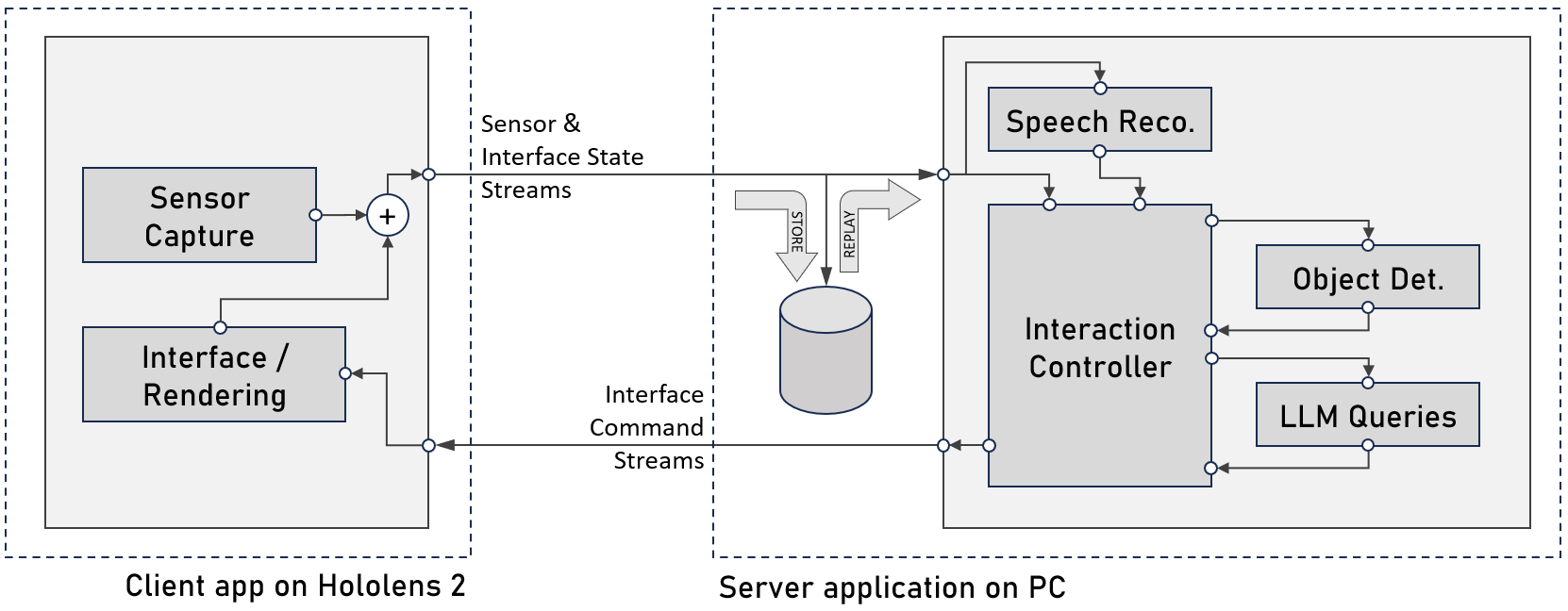}
 \caption{\textsc{Sigma} uses a client-server architecture that bypasses computational limits on device and simplifies development efforts.}
 \label{fig:architecture}
\end{figure*}

\textsc{Sigma} is built using Platform for Situated Intelligence \cite{psi, psi_rapid}, also known as  \textbackslash psi, an open-source framework that provides infrastructure for working with streaming data, a set of components that work with the HoloLens device \cite{psi_mixedreality}, as well as associated development tools. The framework also provides a basic HoloLens stream capture application \cite{psi_hololenscaptureapp} which uses a client-server architecture to capture the various sensor streams from a HoloLens 2 device and transmit them over a network connection to a desktop server computer. 

\textsc{Sigma} adopts and extends the same client-server architecture as the HoloLens stream capture application, illustrated in Figure \ref{fig:architecture}. A \textsc{Sigma} client app runs on device and captures sensor data, including the RGB and depth camera; audio; and head, hand and gaze tracking information. The data is streamed live over a TCP/IP connection to a \textsc{Sigma} server application that runs on a separate desktop computer. The server implements most of the compute functionality and streams back rendering commands to the client app, which uses this information to render UI panels, holograms, and spatial audio accordingly. 

The client-server architecture provides several important affordances for fast prototyping and conducting research. First, the architecture enables researchers to bypass existing computational limits on current headset devices. Pushing the state of the art forward in mixed-reality task assistance requires running live, interactive experimentation with complex models whose GPU/CPU requirements often exceed the computation budget available on current mixed-reality headsets. By offloading the bulk of the computation to the separate server, many possibilities open up. 

\def\approx{\raise.17ex\hbox{$\scriptstyle\sim$}}
\begin{table*}[b]
  \caption{Sensor streams captured by the \textsc{Sigma} client app. In addition to images, the RGB, depth, and preview streams include the corresponding camera intrinsics and extrinsics parameters.}
  \label{tab:sensor_streams}
  \centering%
  \begin{tabu}{lX[l]c}
  \toprule
   Stream & Representation & Frame rate\\
  \midrule
   RGB Camera& $896 \times 504$, NV12 encoding & 5 Hz \\
   Preview Camera & $896 \times 504$, NV12 encoding & 5 Hz \\
   Depth Camera & $320 \times 288$, 16bpp & 5 Hz \\
   Eye Gaze & position ($3 \times 1$ vector) \\ & direction ($3 \times 1$ vector) & \approx30 Hz \\
   Head & pose ($4 \times 4$ matrix) & \approx30 Hz \\
   Hands & poses ($4 \times 4$ matrix) for each of the 26 joints in the left and right hand & \approx20 Hz \\
   Audio & 1-channel, 32-bit floating-point PCM & 16 kHz \\
  \bottomrule
  \end{tabu}%
\end{table*}

Second, the \textsc{Sigma} client-server architecture also enables faster debugging and development by leveraging the data replay capabilities of the underlying \textbackslash psi framework. The server application automatically persists all relevant incoming sensor streams to disk. The existing \textbackslash psi tools can be used to visualize the data and analyze system behavior. The server application can later be re-run directly from the persisted data, eliminating the need to wear the headset and run the live application for each debugging or development session.

Finally, the architecture also fosters device-independent research and development. While in the current implementation the client app targets the HoloLens 2 platform, new versions of the client app could be constructed to run on other devices (\emph{e.g.}, Android phones, Oculus headsets, etc.), while retaining the same server application and features.

\subsection{Client Application}

The \textsc{Sigma} client app captures sensor data to be streamed to the server application, and it receives commands from the server dictating where and how to render graphical elements or spatial sound.

\subsubsection{Sensor capture}

The client app leverages a variety of existing APIs, including HoloLens Research Mode \cite{hololens_researchmode}, to capture the RGB camera stream, the preview stream (i.e., the RGB image with rendered holograms included), the depth camera stream, eye gaze, hand tracking, head pose, and audio from the internal-facing microphone. The various characteristics of these streams, as captured by the current version of the app, are shown in Table \ref{tab:sensor_streams}. 

Given the interactive nature of the system, we made several choices to optimize the sensor capture and minimize latency and on-device compute. For example, while camera images can be captured at higher frame-rates, we opted to sample the camera and preview images at the same clock as the long-range depth images (which are available at 5 Hz). With synchronized depth and RGB data, projection from RGB camera space to 3D point clouds and world coordinates becomes easier and more robust. However, the capture configuration can be easily modified to adjust the resolution and frame-rate for various streams, or to include additional streams such as the 4 on-device gray-scale cameras, accelerometer, magnetometer, and gyroscope data.

\subsubsection{User Interface}

To display holograms and other user interface elements, the \textsc{Sigma} client app leverages StereoKit \cite{stereokit}, an easy-to-use open-source library for building mixed-reality user interfaces with C\# and OpenXR \cite{openxr}. The \textsc{Sigma} user interface currently consists of a floating panel that orients towards the user and displays the recent conversation history along with the current task, step, and sub-step, as shown in Figures \ref{fig:example}.A, B and E. The app also provides support for custom rendering of text, images, 3D models, and timers at locations specified in world coordinates. On first start-up, the application generates and persists a \emph{spatial anchor}\cite{spatialanchors} at the initial pose of the headset in space. That anchor serves as the ``world'' coordinate system that all other poses---such as hand pose, head pose, hologram poses, etc.---are defined in relation to, including in subsequent application runs.

The state of the user interface (\emph{e.g.}, the content of the task and steps, which step is in focus, where the task panel is placed spatially, etc.) is controlled by the server application via a stream of interface commands. Some of the user interface elements can be controlled by the user's hands and gaze. For example, the user can change the location of the task panel by grabbing the gem displayed above it and moving it through space, or simply by saying "come here!" while holding their palm open facing up. Information about the current state of the user interface is passed back to the server on an additional data stream. Together with the affordances of underlying StereoKit \cite{stereokit} and OpenXR \cite{openxr} frameworks, the design patterns provided in the application constitute a basis for easily extending the interface capabilities with other static or actuated mixed-reality user interface elements. 

Finally, the client app can also synthesize speech and render spatial audio. The server application passes information about utterances to be spoken by the system via the interface commands data stream, and speech synthesis is performed via the Azure Speech cloud service \cite{azure_speech}. Events reflecting the moment-by-moment progress of the speech synthesis process are communicated back to the server as part of the interface state data stream.

\subsection{Server Application}

The \textsc{Sigma} server application receives the sensor and interface state data streams from the client app running on device and processes them to control the interaction and to compute rendering commands that are sent back to the client app. The server computation pipeline is centered on the interaction controller component, and relies on additional components for performing speech recognition, tracking relevant objects in the scene, and running large language model queries, as illustrated in Figure \ref{fig:architecture} and explained below.

\subsubsection{Speech Recognition}

\textsc{Sigma} performs speech recognition via a \textbackslash psi component that wraps the continuous speech recognizer from the Azure Speech \cite{azure_speech} service. The component incrementally emits both partial and final recognition results, which are forwarded to the interaction controller. 

\subsubsection{Large Language Model Queries}

\textsc{Sigma} includes a new \textbackslash psi component for making LLM queries that leverages the GPT-4 model available via the Azure OpenAI Service \cite{azure_openai}. The component receives LLM queries from the interaction controller, forwards them to the corresponding cloud service, and provides responses back. A set of APIs support defining and using a library of templated LLM text completion prompts, and enables regression testing. 

\subsubsection{Object Tracking}
\label{sec:object_tracking}

\textsc{Sigma} currently uses off-the-shelf large vision models to detect the objects specified during \emph{gather} steps. Specifically, the system leverages a \textbackslash psi component that can be configured to either use the Detic \cite{detic} or SEEM \cite{seem} zero-shot RGB object detection models. These large vision models can be queried with an image and an open-ended set of objects, and they provide resulting object masks in RGB space.

Objects are detected on a frame-by-frame basis in 2D, and then back-projected to 3D by leveraging the depth sensor (see Figure \ref{fig:example}.D.) This back-projection is accomplished by mapping the 3D points produced by the depth camera to each detection result's pixel masks (using the RGB and depth sensors' intrinsic and extrinsic parameters), resulting in a subset of points for each detected object in 3D space. The centroid of each object's sub-pointcloud is considered as the object's location, and a simple distance-based filtering approach is used to track the detected object instances over time.

\subsubsection{Interaction Controller}
\label{sec:interaction_controller}
The core dialog capabilities of the \textsc{Sigma} system are implemented by an interaction controller component. This component lies at the center of the compute server pipeline and receives the sensor and interface state streams from the client app. It outputs the interface commands stream that is relayed back to the client app, but also a set of command streams for the LLM query and object detection components. In turn, it also receives additional inputs with the results from speech recognition, and LLM query and object detection, as shown in Figure \ref{fig:architecture}.

The interaction controller component maintains an internal representation that captures the state of the interaction (\emph{e.g.}, what is the current step and substep, what is the set of objects that have already been found, etc.). The component updates this state based on incoming inputs, and constructs corresponding interface commands. The current implementation is based on a simple finite-state machine approach with states corresponding to various interaction phases, e.g., identifying the task, providing guidance through a simple or a complex step, etc.

\subsection{Development Tools}
\label{seC:tools}

\begin{figure*}[t]
 \centering 
 \includegraphics[width=\textwidth]{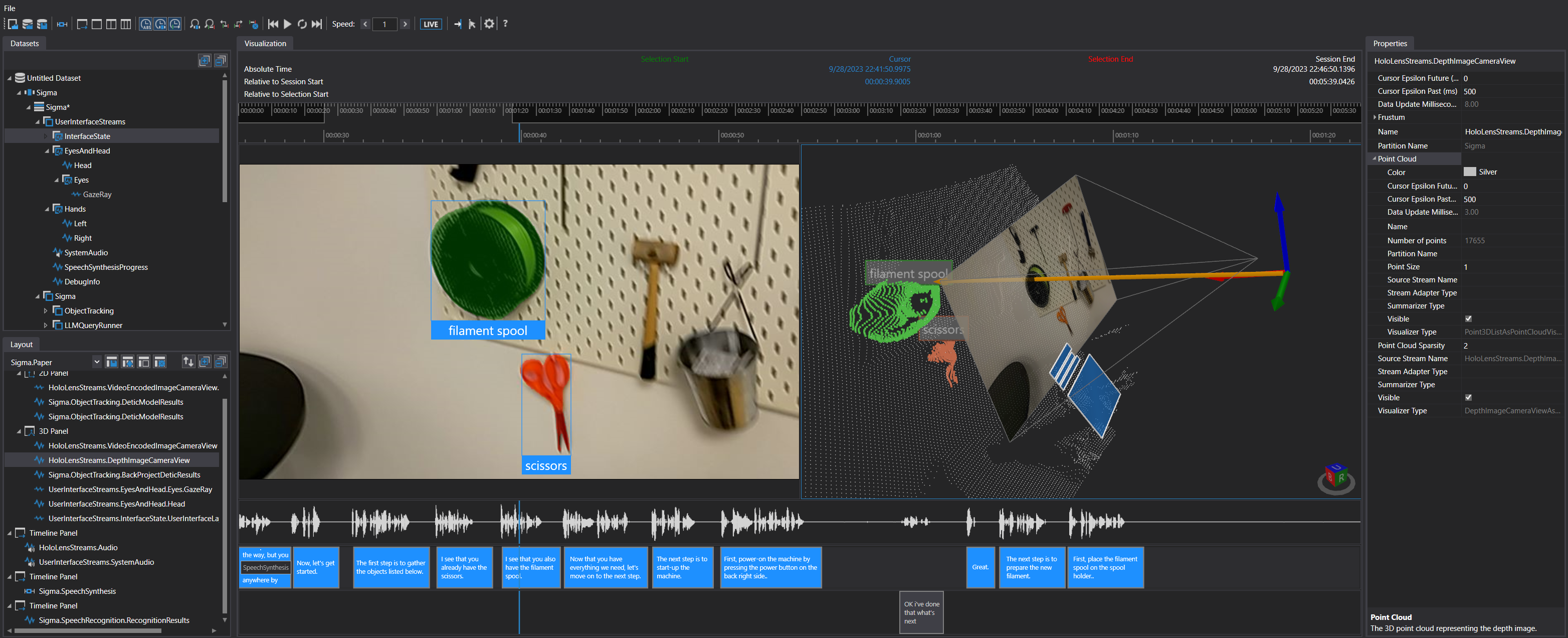}
 \caption{\textsc{Sigma} data visualization constructed using Platform for Situated Intelligence Studio. \textbf{Left panel}: composite 2D visualization containing image and object detection streams. \textbf{Right panel}: composite 3D visualization containing camera image view stream, head coordinate system, gaze direction (orange), UI placement (blue rectangles), depth point cloud, and sub-pointclouds for detected objects. \textbf{Bottom}: timeline visualizations for audio, speech synthesis and speech recognition result streams.}
 \label{fig:psi_studio}
\end{figure*}

Researchers using \textsc{Sigma} can leverage and benefit from the tools and affordances of the underlying \textbackslash psi framework. \textbackslash psi provides an efficient infrastructure for data logging, and all the data streams captured or generated by the application can be automatically persisted to disk. The collected data can be easily inspected (offline or live while the system is running) using Platform for Situated Intelligence Studio \cite{psi_rapid}---a tool that allows for visualizing and navigating over the streaming data. For example, Figure \ref{fig:psi_studio} shows a custom visualization layout for \textsc{Sigma} that includes the camera view with overlaid Detic object detection masks; the 3D world view with the user's head pose, gaze direction, RGB camera frustum, depth camera point cloud, interface panels, and back-projected object detections; and captured audio, speech recognition, and speech synthesis streams. Overall, the tool simplifies development, debugging and maintenance efforts by allowing developers to create custom visualization layouts, navigate over and inspect temporal data streams, create temporal manual annotations, and run batch processing tasks \cite{psi, psi_rapid}. In addition, \textsc{Sigma} provides a utility tool that allows researchers to manually place holograms in the world and extract corresponding poses expressed in world-coordinates to include in the task library.

\section{Discussion}
\label{sec:Discussion}

\textsc{Sigma} provides an initial implementation of an end-to-end mixed-reality procedural task assistance system. The set of features and the functionality implemented by this initial version are not meant to be comprehensive. Rather, they are meant to provide a starting point for exploration at the intersection of mixed reality and AI. In this section, we briefly review some of the lessons learned from the development of this baseline system, and discuss possible extensions and research directions. 

Early in the system's development, we made the choice to bypass the computational limits of the HoloLens 2 device and pursue a distributed architecture that allows for more compute-intensive models to run on a separate compute server. This choice carries clear benefits but also challenges. Minimizing the compute requirements on the device allows the app run more smoothly, but live-streaming the sensor data also creates  demands and trade-offs in terms of networking throughput, latency and reliability. Reasoning about latency and synchronizing streams as they flow over the network and through the server computation pipeline (different models operate at different speeds) is a critical need that was well supported by the runtime and libraries provided by \textbackslash psi. Another benefit of this architectural choice is that it opens the door to more easily swapping out devices for the client app while leaving the server side mostly unchanged.

\textsc{Sigma}'s modular design makes it a promising platform for further prototyping and research. Researchers can easily swap out components to improve certain capabilities, or to introduce and test new models in the context of an existing end-to-end application. For example, while the system originally relied on the Detic \cite{detic} object detection model, we later enabled it to use the alternative panoptic segmentation from SEEM \cite{seem}. Similarly, we envision other researchers might replace the cloud-based Azure Speech recognition component with an open-source multilingual model like Whisper \cite{whisper}, or introduce and use additional multimodal foundation models such as LLaVA \cite{llava}, Kosmos \cite{kosmos} or LAVIS \cite {lavis}. Researchers could even integrate their own custom models for scene understanding, event detection, action recognition, etc.

The user interface driving the system interaction requires significant usability testing and there are undoubtedly many opportunities for improvement. For example, one salient question that arose early in development was where the interface panels (including the task instructions and chat bubbles) should be placed in 3D space. In the current design, the interface starts in front of the user. Once the task is identified, it moves itself to the side to get out of the way. While the user can manually adjust where the interface is placed anytime throughout the task, the UI can end up outside the field of view, intersect with other objects, or occlude the task space. Instead, an intelligent interface might reason about the active workspace and relevant objects in it, as well as the user's state, \emph{e.g.}, their location and focus of attention over time, and automatically place itself in accessible positions without occluding the task at hand.

Expanding the interaction capabilities of the system is another important area of future work and a source of open-ended research questions. For instance, future extensions of the system should support more complex task recipes, including branching and preconditions. Advanced capabilities will enhance the usability of the system by enabling it to be more proactive and automatically advance to the next step, intervene with advice or additional details when necessary, and so forth.

The use of LLMs for procedural task assistance is also an area deserving additional design and research focus. The current prompts used by \textsc{Sigma} were tuned in successive iterations. However, the open-ended and stochastic nature of these generative models makes them difficult to control: the same prompt is not guaranteed to produce the same response when queried multiple times, and switching from one LLM to another requires attention to and sometimes redesigning the prompts. Much remains to be done in increasing LLM robustness, particularly when generating task content that is aligned with the user's context. Other research directions we plan to investigate moving forward include the use of multimodal foundation models \cite{kosmos, llava, openflamingo, seem, lavis} to better understand user context and the actions being performed, and compositing information from different models together through Socratic model approaches \cite{socratic}.

More broadly, \textsc{Sigma} is also intended to serve as a platform for research on developing new models and technologies in the space of human-AI collaboration. It can serve as a testbed for conducting research on scene understanding of 3D objects and environments that evolve dynamically over time; human cognitive state estimation, \emph{e.g.}, understanding when the user might be frustrated, confused, distracted, or in a state of flow, etc.; action and activity recognition, \emph{e.g.}, understanding when a user has completed a step and whether the step was performed correctly; research on incremental dialog interleaved with physical actions to enable fluid, seamless collaboration between users and the system; and so on.

As researchers in machine learning, computer vision, natural language processing, etc., develop new models supporting these and other capabilities, it will be important to test them in the context of complete end-to-end interactive systems, rather than just static, dataset-based benchmarks. We believe \textsc{Sigma} can be used to help address that gap, and provide a valuable empirical testbed for conducting model evaluations in an interactive setting. 

\section{Responsible AI Considerations}

\textsc{Sigma} was designed as an experimental prototype for research purposes only, and is not intended for use in developing commercial applications. The primary use case is as a research tool to enable academic and industry researchers to push the state of the art in the space of procedural task assistance at the intersection of mixed reality and AI. As such, the system has been open-sourced under a research-only license\footnote{\url{https://github.com/microsoft/psi/blob/master/LICENSE-APPLICATIONS.txt}}.

In addition to business or mission critical purposes, \textsc{Sigma} should also not be used for any medical or health-related tasks, or for tasks that are dangerous or have an increased risk of physical harm.

\textsc{Sigma} relies on the HoloLens 2 headset, LLMs, and off-the-shelf vision models, all of which contribute various risks and limitations to the system. Wearing the HoloLens 2 headset for long periods of time may lead to discomfort, and the headset (including virtual content it is rendering) may sometimes visually occlude physical objects in the environment. Users must therefore carry out tasks with extra care and caution. LLMs can be configured to generate recipes and answer questions, but LLMs are known to exhibit data biases, lack of contextual understanding, lack of transparency, potential for inaccurate content, and so on. In addition, vision models still have many accuracy and robustness issues today, and may inaccurately classify objects in the environment, leading to incorrect answers, missing context, and inappropriate instructions. Sometimes these inaccuracies might lead users into unsafe situations, which is why they must cautiously exercise their own best judgment at all times.

Researchers that wish to make use of \textsc{Sigma} in their own work should first familiarize themselves with the system and its limitations and risks involved with using the system in a user-study context. When configuring SIGMA to use LLMs, we recommend testing during early pilot experimentation that the responses provided by the models are reasonable for the tasks and domains being investigated. Researchers that wish to conduct studies with \textsc{Sigma} and human participants should undergo a full IRB or ethical board review as appropriate for their institution. As part of any user study protocol, researchers should communicate to the participants all known limitations and the risks of using the system, as well as the data types collected.

All of these considerations and more are detailed in a Transparency Note\footnote{\url{https://github.com/microsoft/psi/wiki/Sigma-Transparency-Note}} available in \textsc{Sigma}'s open-source repository.

\section{Conclusion}

We have introduced \textsc{Sigma}, an open-source end-to-end interactive system for mixed-reality procedural task assistance. The system harnesses the multimodal sensing and rendering capabilities of the HoloLens 2 device, together with large language and vision models to engage in mixed-initiative dialog and guide users step by step through procedural tasks. The system is built on a modern infrastructure with associated tools for working with temporally streaming data, and has a modular design that fosters research and exploration. We hope to engage the research community and together extend its capabilities and use it as a springboard for new avenues of research.

\footnotesize
\section*{Acknowledgements}
We would like to thank Neel Joshi, Vibhav Vineet and Xin Wang for useful discussions and comments during the development of the system.

\normalsize
\bibliographystyle{abbrv-doi}
\bibliography{main}


\end{multicols}  
\end{document}